\documentclass[journal]{IEEEtran}
\usepackage{algorithm}
\usepackage{algorithmicx}
\usepackage[noend]{algpseudocode} 
\usepackage[colorlinks=true, allcolors=blue]{hyperref}
\usepackage{graphicx}
\usepackage{float}
\usepackage{amsmath}
\usepackage{amsmath,amsfonts,amssymb}
\usepackage{graphicx}
\usepackage{setspace}
\usepackage{tocloft}
\usepackage{textcomp}
\usepackage{float}
\usepackage{amsmath,amsfonts,amssymb}
\usepackage{graphicx}
\usepackage{multirow}
\usepackage[table,xcdraw]{xcolor}
\usepackage{lineno}
\newcommand{\Fig}{Fig.}

\newcommand{\Tab}{Tab.}

\begin{document}
%
\title{Instance Segmentation for Whole Slide Imaging:\\ End-to-End or Detect-Then-Segment}
%
%
%

\author{Aadarsh~Jha,
        Haichun~Yang,
        Ruining~Deng,
        Meghan~E.~Kapp,
        Agnes~B.~Fogo,
        and~Yuankai~Huo,~\IEEEmembership{Member,~IEEE}
\thanks{A. Jha, R. Deng,  Y. Huo were with the Department of Electrical Engineering and Computer Science, Vanderbilt University, Nashville,
TN, 37215, USA, e-mail: yuankai.huo@vanderbilt.edu}
\thanks{H. Yang, M. Kapp, A. Fogo were with the Department
of Pathology, Vanderbilt University Medical Center, Nashville,
TN, 37215, USA}
}

\markboth{Manuscript pre-print, July~2020}%
{Shell \MakeLowercase{\textit{et al.}}: Bare Demo of IEEEtran.cls for IEEE Journals}

\maketitle

\begin{abstract} Automatic instance segmentation of glomeruli within kidney Whole Slide Imaging (WSI) is essential for clinical research in renal pathology. In computer vision, the end-to-end instance segmentation methods (e.g., Mask-RCNN) have shown their advantages relative to detect-then-segment approaches by performing complementary detection and segmentation tasks simultaneously. As a result, the end-to-end Mask-RCNN approach has been the de facto standard method in recent glomerular segmentation studies, where downsampling and patch-based techniques are used to properly evaluate the high resolution images from WSI (e.g., $>$ 10,000$\times$10,000 pixels on 40$\times$). However, in high resolution WSI, a single glomerulus itself can be more than 1,000$\times$1,000 pixels in original resolution which yields significant information loss when the corresponding features maps are downsampled to the 28$\times$28 resolution via the end-to-end Mask-RCNN pipeline. In this paper, we assess if the end-to-end instance segmentation framework is optimal for high-resolution WSI objects by comparing Mask-RCNN with our proposed detect-then-segment framework. Beyond such a comparison, we also comprehensively evaluate the performance of our detect-then-segment pipeline through: 1) two of the most prevalent segmentation backbones (U-Net and DeepLab\textunderscore v3); 2) six different image resolutions (512$\times$512, 256$\times$256, 128$\times$128, 64$\times$64, 32$\times$32, and 28$\times$28); and 3) two different color spaces (RGB and LAB). Our detect-then-segment pipeline, with the DeepLab\textunderscore v3 segmentation framework operating on previously detected glomeruli of 512$\times$512 resolution, achieved a 0.953 dice similarity coefficient (DSC), compared with a 0.902 DSC from the end-to-end Mask-RCNN pipeline. Further, we found that neither RGB nor LAB color spaces yield better performance when compared against each other in the context of a detect-then-segment framework. Detect-then-segment pipeline achieved better segmentation performance compared with End-to-end method. This study provides an extensive quantitative reference for other researchers to select the optimized and most accurate segmentation approach for glomeruli, or other biological objects of similar character, on high-resolution WSI.
\end{abstract}

\begin{IEEEkeywords}
Segmentation, Deep Learning, U-Net, Mask-RCNN, Glomeruli, Whole Slide Imaging
\end{IEEEkeywords}

%
\IEEEpeerreviewmaketitle

\section{Introduction}
\begin{figure*}[t]
\begin{center}
\includegraphics[width=1.0\textwidth]{{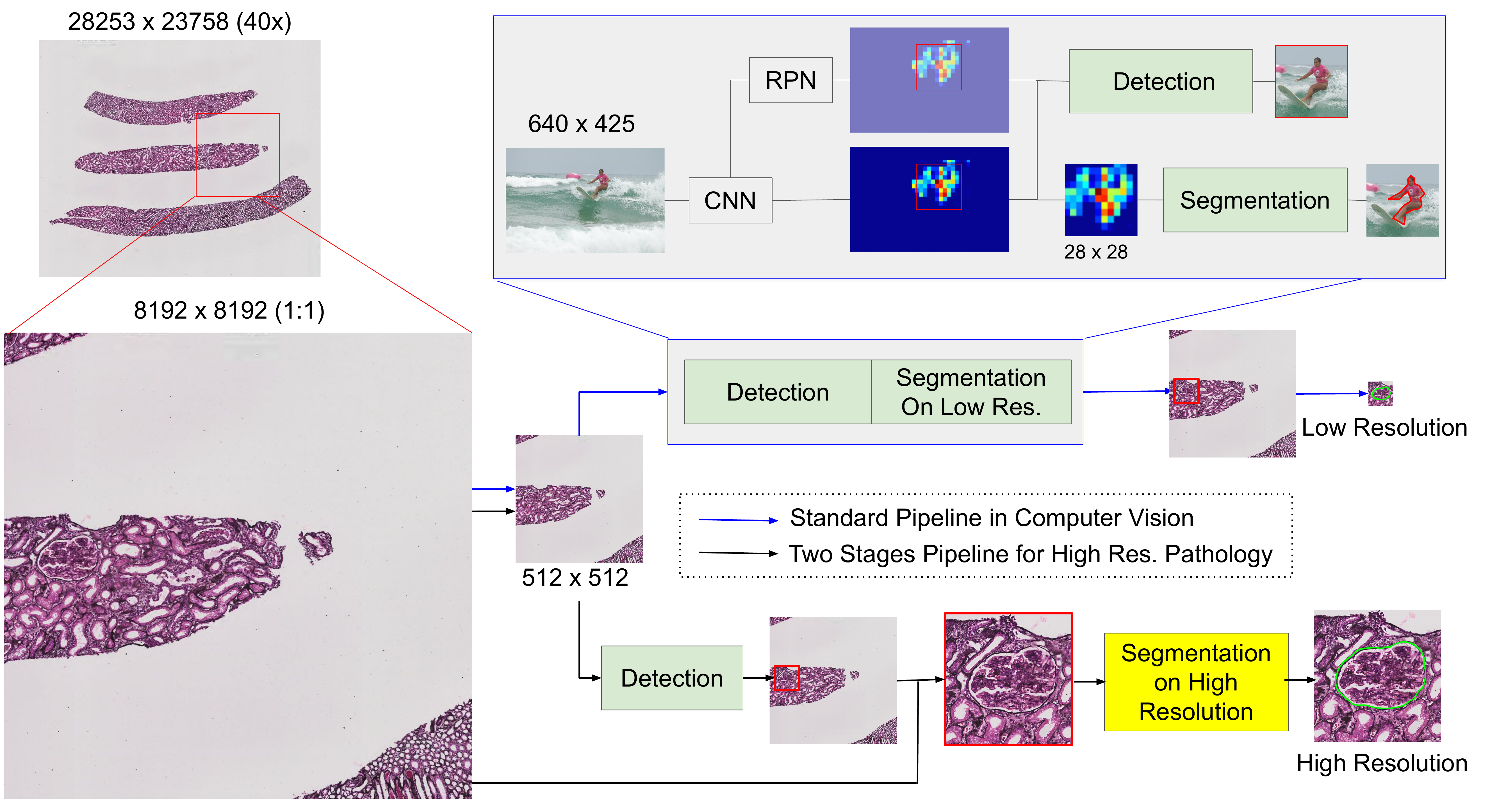}}
\end{center}
\caption{
  This figure showcases the end-to-end Mask-RCNN instance segmentation pipeline in blue arrows, and the proposed detect-then-segment framework in black arrows. In our proposed method, the two-stage detect-then-segment strategy is used, where detection will first occur on downsampled images, and then segmentation is performed on high resolution objects.
} 
\label{Fig.1} 
\end{figure*}

\IEEEPARstart{U}{nderstanding} the underlying details of glomerular morphology through renal biopsy evaluation provides insights into various renal disorders~\cite{greenberg2009primer,rayat2007glomerular,puelles2015counting}. The golden standard of characterizing glomeruli is a manual estimation via advanced imaging techniques~\cite{puelles2015counting}. However, manual quantification of morphometric parameters requires exhaustive resources and is not scalable. In recent years, there has been a paradigm shift towards automatic glomerular instance segmentation, which aims to provide instance-level, pixel-wise annotation for each glomerulus driven by Convolutional Neural Networks (CNNs)~\cite{gallego2018glomerulus, bukowy2018region}. The de facto standard method of instance segmentation of glomeruli, and more broadly the kidney, is Mask-RCNN~\cite{peng2019extent,macdonald2020improved}, an end-to-end pipeline which performs detection and instance segmentation simultaneously~\cite{he2017mask}. Since the end-to-end architecture of Mask-RCNN is designed for natural images (e.g., $\approx$ 1000$\times$1000 pixels), both downsampling and tiling are utilized in order to leverage processing speeds and fit modern GPU memory when Mask-RCNN is applied to high resolution WSI (e.g., $>$ 10000$\times$10000 pixels on 40x). However, a loss of information is often associated with the downsampling process that is inherent to the end-to-end framework of Mask-RCNN. In particular, a single glomerulus from a WSI can be more than 1,000$\times$1,000 pixels in image resolution, which yields significant information loss when the corresponding features maps are downsampled to the 28$\times$28 resolution via the end-to-end Mask-RCNN segmentation head~\cite{peng2019extent}, as demonstrated in {\color{red} \Fig~\ref{Fig.1}}. Thus, the prevalent end-to-end instance segmentation method might not be the best solution for high resolution WSI. When re-imagining glomerular instance segmentation for high resolution WSI, an intuitive idea of addressing the tradeoff between resolution and accuracy would be the fundamental separation of detection and segmentation. In this detect-then-segment manner, detection could be performed on downsampled tiles for computational efficiency, while segmentation could be conducted on high-resolution images as unrelated pixels are excluded by detection. Inspired by this rationale, we aim to explore if the end-to-end or detect-then-segment framework is optimal for high-resolution WSI objects in the context of renal pathology. 

In the current study, we propose a detect-then-segment framework for glomerular instance segmentation in order to more broadly improve current instance segmentation techniques when applied to high-resolution WSI. In our study, we utilize two distinct high-resolution segmentation networks for semantic segmentation, and we use Mask-RCNN for instance glomerular detection. A central focus of our study is to compare our proposed detect-then-segment framework to the performance of the end-to-end Mask-RCNN pipeline on high resolution WSI. In addition, we conduct extensive analyses to ascertain the best detect-then-segment strategy through two of the most widely used segmentation backbones (U-Net and DeepLab\textunderscore v3), six unique resolutions (512$\times$512, 256$\times$256, 128$\times$128, 64$\times$64, 32$\times$32, and 28$\times$28), and two distinct color spaces (RGB and LAB). To the best of our knowledge, no previous studies have comprehensively evaluated glomerular segmentation performance comparing detect-then-segment and end-to-end strategies. 

To evaluate the performance of these two distinct segmentation frameworks, we divided our experiments into two scenarios: 1) manual detection, and 2) automatic detection. In the first scenario --- labeled as ``manual detection" --- manual detection results (bounding boxes) were used to evaluate the segmentation performance in our detect-then-segment framework. Then, in the second scenario --- labeled as ``automatic detection" --- automatic detection results from the same Mask-RCNN detection head were used to compare segmentation performance across end-to-end and detect-then-segment strategies. The key difference in our automatic detection phase is the use of either an end-to-end Mask-RCNN segmentation head or an additional high-resolution segmentation head for glomerular instance-level segmentation. Decoupling detection and segmentation allows for more freedom in understanding how to improve the segmentation of glomeruli, and more broadly, high resolution WSI objects of similar character.

For the manual detection phase, we trained the segmentation networks of our proposed detect-then-segment method using 704 manually traced training glomerular images from 42 biopsy samples. Meanwhile, 98 validation glomerular images, 147 internal testing images, and 385 external testing images were manually extracted from 7, 7, and 5 WSI images respectively to evaluate segmentation performance. The original resolution of our glomeruli image data was of $\approx$ 1000$\times$1000 pixels. To be compatible with our GPU memory, we first scaled all input images down to the resolution of 512$\times$512. Then, according to our procedure, we further scaled down the input images to the resolutions of 512$\times$512, 256$\times$256, 128$\times$128, 64$\times$64, 32$\times$32, and 28$\times$28 to train and evaluate two of the most widely used segmentation backbones, U-Net and DeepLab\textunderscore v3. 

For the automatic detection experiments, we compared performance between the end-to-end Mask-RCNN segmentation and our proposed detect-then-segment approach using automatic detection results. To do so, Mask-RCNN was trained and validated on the same biopsy WSI images as the manual detection experiments. From the 4 internal testing WSI, 120 glomeruli were correctly detected (IOU $>$ 0.5 compared with true bounding box) from the detection head in Mask-RCNN. Then, we directly applied the trained Mask-RCNN segmentation head as well as our trained segmentation models (U-Net and DeepLab\textunderscore v3) from our first phase to the same 120 glomerular detected images to compute the final segmentation performance when the detection is fairly provided. The experiments show that our detect-then-segment framework, under the automatic detection scenarios, achieves a DSC value of 0.953, whereas Mask-RCNN provides a lower DSC value of 0.902. 

Our contributions, as listed below, do not claim algorithmic novelty over prior arts but rather investigate the problems overlooked in previous works:
 \begin{itemize}
  \item Proposing a new detect-then-segment glomerular instance segmentation framework by performing instance detection and semantic segmentation on different resolutions with a coarse-to-fine design to avoid extreme downsampling for high resolution glomerular segmentation in renal pathology WSI.  
  \item Evaluating if the de facto end-to-end design or our detect-then-segment approach is optimal for segmenting glomeruli in high resolution WSI.
  \item Performing extensive analyses by varying the image resolution, color space, and segmentation framework in segmenting previously detected glomeruli image objects. This comprehensive analysis allows us to provide an extensive quantitative reference for other researchers to select the optimized segmentation approaches for glomeruli, or other biological objects of similar character, on high-resolution WSI.
\end{itemize}

\begin{figure*}[t]
\centering 
\includegraphics[width=1\textwidth]{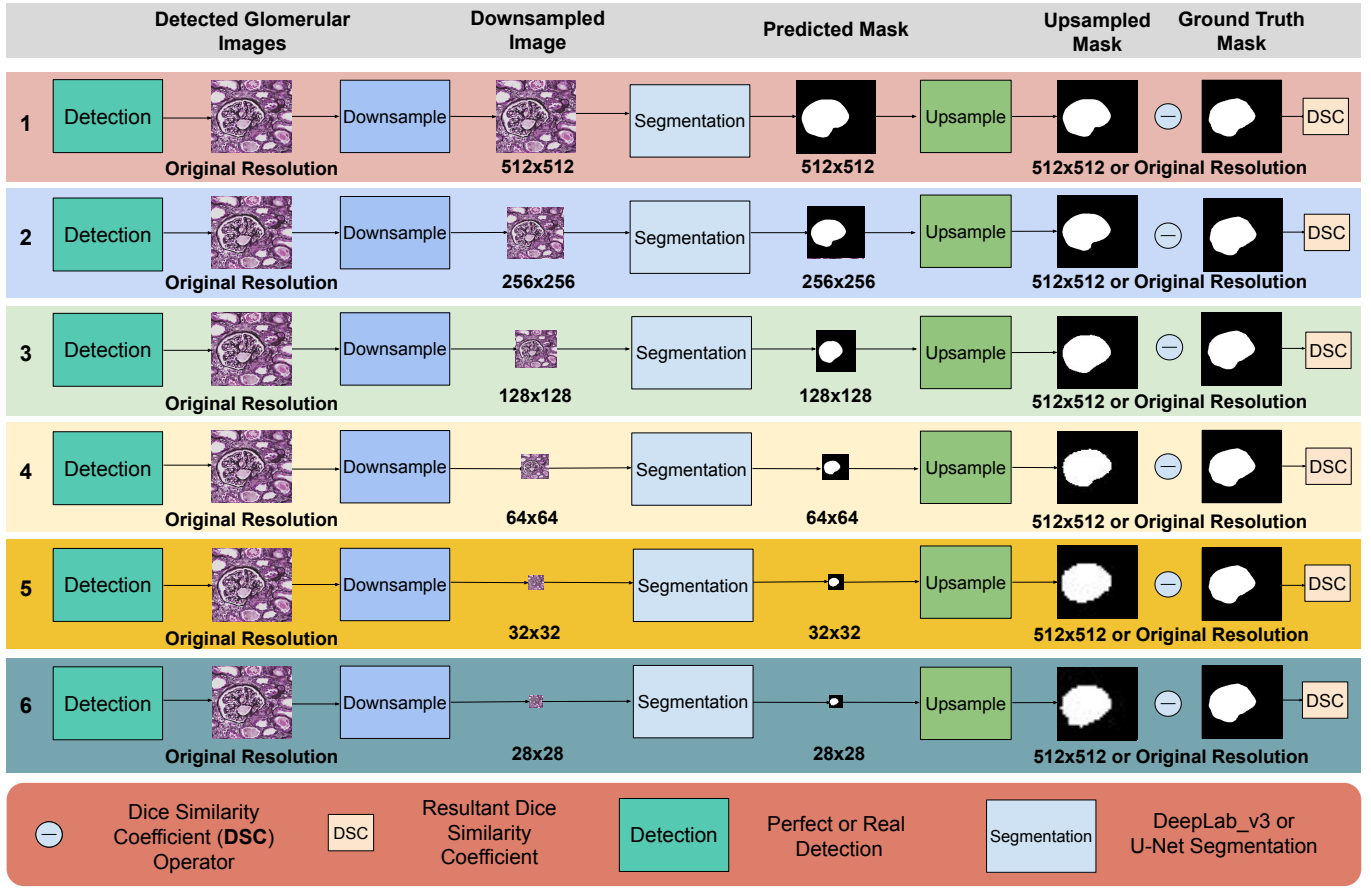}
\caption{
   This figure demonstrates an abstraction of the proposed detect-then-segment methodology used in our experiments. Each row indicates a different trial, where a previously detected glomerulus is downsampled to the distinct dimensions of 512$\times$512, 256$\times$256, 128$\times$128, 64$\times$64, 32$\times$32, and 28$\times$28. Then, these downsampled images are passed through a segmentation network. Two of the most prevalent segmentation backbones (U-Net and Deeplab\textunderscore v3) are used as the segmentation networks in this study. The predicted masks are produced, and then upsampled back to the initial and original resolution --- which in our study was 512$\times$512 --- of the glomerulus for a fair DSC comparison. Additionally, in the trial utilizing a U-Net backbone, we separately evaluated the images in both a RGB and LAB color space so as to understand the effect of color space on segmentation performance. In the Deeplab\textunderscore v3 trial, we only evaluated the best performing color space from our U-Net experiment. Overall, this figure shows the segmentation networks evaluated across six different resolutions, two unique color spaces, and two distinct segmentation backbones.
} 
\label{Fig.2} 
\end{figure*}

\section{Related Works}
The introduction of WSI demonstrates a shift towards computer-aided diagnosis (CAD) techniques to more accurately characterize critical objects. The use of WSI, and its associated analysis techniques, has been shown to be effective and even expanding in the field of renal pathology~\cite{sarwar2019physician}. To properly distinguish and characterize different glomeruli within renal biopsy samples, modern deep learning techniques of detection and segmentation have been utilized. Several studies have shown the great accuracy by which CNNs are able to properly detect and localize glomeruli within sample images~\cite{bukowy2018region, uchino2020classification, pavinkurve2019deep, temerinac2017detection}.  Similarly, CNNs have also been able to accurately segment glomeruli, allowing for normal and sclerosed glomeruli to be properly distinguished~\cite{altini2020semantic, kannan2019segmentation, gadermayr2017segmenting, gadermayr2017cnn, ginley2019computational, ginley2020neural}. Other studies have further combined the process of the detection and segmentation of glomeruli~\cite{bueno2020glomerulosclerosis}. Of course, the end-goal of deep learning in renal imaging is its application in CAD. In this regard, several studies have also shown the ability to perform diagnoses based on the preliminary quantification and characterization of glomerular data through deep learning~\cite{barros2017pathospotter,marsh2018deep}. Common to all of the above studies is the application of CNNs to localize, detect, segment, or characterize glomeruli to better understand renal pathology. The uniqueness of our research presents itself by identifying the specific techniques that work best for high resolution glomeruli data, rather than using common solutions to the niche field of renal imaging. Our paper analyzes several factors that work best in the specific context of instance segmentation of high resolution glomerular data, and other biological objects of similar character and size. Additionally, we further propose a pipeline that is different than the conventional end-to-end instance segmentation tactics that are often used in medical imaging and computer vision, so as to yield better and more accurate results. 

\section{Method}

Generally, the methodology followed in this study can be broken down into two major steps: 1) detection and 2) segmentation. Within detection, we discuss our approach towards manual and automatic detection of glomeruli; on the other hand, within segmentation, we demonstrate how we comprehensively analyzed our detect-then-segment framework, and the steps taken to compare it to a classic end-to-end Mask-RCNN pipeline. Our detect-then-segment approach can be seen visually in {\color{red} \Fig~\ref{Fig.2}}.

 \subsection{Detection}
 
In the manual detection portion of our study, the manually traced bounding boxes for glomeruli are used to provide the ideal detection results. In order to introduce more background variation and avoid the problematic situation in which the glomeruli is always in the middle of the detection, we randomly expanded the detection bounding boxes to 1.5 times the longest dimension of the manual boxes with random center shift. We ensured that the image still contained the complete glomerulus. 

In the automatic detection portion, Mask-RCNN~\cite{he2017mask} was employed as the detection method. The Feature Pyramid Network (FPN)~\cite{lin2017feature} with ResNet-101~\cite{he2016deep} is used as the feature extraction backbone. The default Mask-RCNN implementation (\url{https://github.com/facebookresearch/maskrcnn-benchmark}) was used during training. For all training and testing within detection, the original high-resolution WSI (0.25 $\mu m$ per pixel) was downsampled to a lower resolution (4 $\mu m$ per pixel), given the size of a glomerulus~\cite{puelles2011glomerular} as well as its ratio within a patch. Then, we randomly tiled the image patches (where each patch contained at least one glomerulus with 512$\times$512 pixels) as experimental images for our detection networks. Eventually, we formed a cohort with 7040 training images with manual segmentation masks for training the Mask-RCNN glomerular detection.

\subsection{Segmentation}

\subsubsection{Manual Detection}
In this study, a standard implementation of the U-Net and DeepLab\textunderscore v3 architectures was used to perform segmentation on the glomeruli image data in the manual detection phase of our experiment. In particular, U-Net and DeepLab\textunderscore v3 were trained with the preprocessed images as described in \textsection{3.1}. The input image data for both segmentation frameworks contained 3 input channels (RGB or LAB), and the output data contained 2 classes (foreground and background). 

Limited by GPU memory, all original image resolution glomeruli ($>$ 1000$\times$1000) were initially scaled down to 512$\times$512. Then, this input image dataset was further scaled down to the sizes of 512$\times$512, 256$\times$256, 128$\times$128, 64$\times$64, 32$\times$32, and 28$\times$28. Once these images were downsampled, the training images were further represented in either the standard RGB or LAB image space. The LAB image space was evaluated as it was recently shown to confer the best performance for basic image classification tasks by reducing image channel-wise correlation~\cite{gowda2018colornet}. Data augmentation was also performed for image segmentation, where 50\% of the training images were altered through channel shuffling, translation, rotation, sheer, left-to-right flipping, and Gaussian blur.

Two of the most prevalent segmentation backbones (U-Net and DeepLab\textunderscore v3) were employed in this study. Briefly, the U-Net architecture is an end-to-end Fully Convolutional Network (FCN). In terms of general network architectures, U-Net can be divided into two major portions: 1) encoder, and 2) decoder. The encoder contains both convolutional and max pooling layers which obtain greater context of the input image through downsampling, allowing for the encoding of the input image into feature representations at multiple different resolutions. The second path is the decoder, which symmetrically expands and upsamples the input image. This allows for precise localization using bilinear interpolation, and effectively rescales the feature map to the original image size~\cite{ronneberger2015u}. Similarly, DeepLab\textunderscore v3 also has encoder and decoder stages. However, in its encoder phase, DeepLab\textunderscore v3 utilizes Atrous, or dilated, convolution to obtain greater context of the input image. The decoder phase then follows to create and rescale the feature map of the original image~\cite{he2017mask}. Through the manually detected glomerular images, we evaluated the performance of U-Net and DeepLab\textunderscore v3 with the aforementioned designs.

\subsubsection{Automatic Detection}

Finally, in the automatic detection phase of our experiment, the trained Mask-RCNN network was performed on all testing images to achieve 120 glomerular detection bounding boxes from 5 WSI biopsies in downsampled images. Then, the bounding box coordinates were upscaled to the original image resolution ($>$ 1000$\times$1000 pixels) to crop the corresponding glomeruli and the masks in highest resolution. Furthermore, both DeepLab\textunderscore v3 and U-Net pretrained models were also applied on the same group of 120 images, which were downscaled to each of the tested resolutions (512$\times$512, 256$\times$256, 128$\times$128, 64$\times$64, 32$\times$32, and 28$\times$28). After corresponding predicted masks were generated, they were upsampled to the initial resolution to calculate the mean and median DSC scores from the manual masks.

\subsection{Data Analysis}

The DSC was primarily used to evaluate the performance of segmentation. To begin, in our manual detection experimentation --- which comprised of the 704 training, 98 validation, 147 internal testing, and 385 external testing images --- we evaluated the performance of U-Net and DeepLab\textunderscore v3 segmentation across six different resolutions (512$\times$512, 256$\times$256, 128x129, 64$\times$64, 32$\times$32, 28$\times$28) and two different color spaces (RGB and LAB). In particular, for each epoch within the segmentation process for each resolution, DSC values were developed for the validation and testing data. For each tested resolution, the best epoch was selected via the highest DSC for the validation dataset, and the generated model in that epoch was saved. Then, these generated, predicted masks for each tested resolution were upsampled to a 512$\times$512 resolution to be compared against the initial ground truth mask data (which is also of 512$\times$512 resolution) for the validation and testing images. Mean and median DSC, as well as standard deviation, were computed again for these upsampled image sets for each resolution.

Throughout our study, in the specific context of our manual detection phase results, we draw a distinction between the terms of ``sample space" and ``512 space". We define ``sample space" as the evaluation of the predicted images in each of the six tested resolutions (512$\times$512, 256$\times$256, 128$\times$128, 64$\times$64, 32$\times$32, and 28$\times$28) against the corresponding downsampled input images to produce a preliminary DSC value across six distinct resolutions. We similarly define ``512 space" as the evaluation of the predicted images across the six tested resolutions which are then upsampled to the original size of the input image --- which in our study was 512$\times$512 as established earlier --- and then compared to the original resolution 512$\times$512 input images to produce a fair DSC score. This can be seen visually in the latter columns of {\color{red} \Fig~\ref{Fig.2}}. 

Furthermore, in our automatic detection experimentation, which comprised a cohort of 120 glomerular images in original resolution ($>$ 1000$\times$1000), we similarly applied the detect-then-segment framework which was directly compared to Mask-RCNN through the use of mean and median DSC, as well as the standard deviation of the DSC data. Through a similar process in the manual detection phase, we applied Mask-RCNN to our cohort of input images, and produced relevant statistics for the DSC scores. Additionally, after applying both U-Net and DeepLab\textunderscore v3 to each of the six tested resolutions of the input image and producing corresponding predicted masks, such predicted masks were then upsampled and compared to the original resolution of the input image ($>$ 1000$\times$1000) for a fair DSC comparison.

\section{Experimental Design}

\subsection{Dataset}
WSI from renal needle biopsies and human kidney nephrectomy tissues were utilized for analysis. The kidney needle biopsy was routinely processed, paraffin embedded, and 2 $\mu m$ thickness sections cut and stained with hematoxylin and eosin (HE), periodic acid–-Schiff (PAS) or Jones. The human nephrectomy tissues were acquired from noncancerous tissue from patients with cancer. The tissue was routinely processed, paraffin embedded, and 3 $\mu m$ thickness sections cut and stained with PAS. The data was deidentified, and studies were approved by the Institutional Review Board (IRB). For the purposes of training and testing, the high resolution WSI (0.25 $\mu m$ per pixel) was downsampled to a lower resolution (4 $\mu m$ per pixel). Then, patches were identified which contained glomeruli in its original resolution ($>$ 1000$\times$1000 pixels). Images of glomeruli, as well as their manually traced ground truth masks, were then collected. In this study, these input images served as our previously detected glomerular images upon which segmentation then was performed. Eventually, we formed a cohort with 704 training, 98 validation, and 147 internal testing images. Additionally, a group of 385 images was used as external testing data. The training, validation, and testing data were used in our manual detection experimentation. Finally, a separate cohort of 120 images with Mask-RCNN detected glomeruli was used to directly evaluate the performance of our proposed framework relative to Mask-RCNN. This set of 120 images was derived from 5 different patients with WSI of the kidney tissue, and was utilized in our automatic detection experimentation. 

\begin{table*}
\caption{DSC scores collected for the internal testing data using both U-Net and DeepLab\textunderscore v3 in 512 Space.}

\centering
\begin{tabular}{|c|c|l|l|l|l|l|l|}
\hline
\multicolumn{2}{|c|}{Image Resolution}                                                        & \multicolumn{1}{c|}{512$\times$512}                             & \multicolumn{1}{c|}{256$\times$256}                            & \multicolumn{1}{c|}{128$\times$128}                            & \multicolumn{1}{c|}{64$\times$64}                              & \multicolumn{1}{c|}{32$\times$32}                              & \multicolumn{1}{c|}{28$\times$28}                              \\ \hline
\multirow{2}{*}{U-Net}       & \begin{tabular}[c]{@{}c@{}}Mean Dice \\ $\pm$ Std Dev\end{tabular} & \begin{tabular}[c]{@{}l@{}}0.909 \\ $\pm$ 0.099\end{tabular} & \begin{tabular}[c]{@{}l@{}}0.936\\ $\pm$ 0.073\end{tabular} & \begin{tabular}[c]{@{}l@{}}\textbf{0.940}\\ \textbf{$\pm$ 0.051}\end{tabular} & \begin{tabular}[c]{@{}l@{}}0.920\\ $\pm$ 0.047\end{tabular} & \begin{tabular}[c]{@{}l@{}}0.878\\ $\pm$ 0.066\end{tabular} & \begin{tabular}[c]{@{}l@{}}0.853\\ $\pm$ 0.072\end{tabular} \\ \cline{2-8} 
                             & Median Dice                                                    & 0.948                                                    & \textbf{0.961}                                                   & 0.957                                                   & 0.936                                                   & 0.897                                                   & 0.872                                                   \\ \hline
\multirow{2}{*}{DeepLab\_v3} & \begin{tabular}[c]{@{}c@{}}Mean Dice \\ $\pm$ Std Dev\end{tabular} & \begin{tabular}[c]{@{}l@{}}\textbf{0.948}\\ \textbf{$\pm$} \textbf{0.062}\end{tabular}  & \begin{tabular}[c]{@{}l@{}}0.947\\ $\pm$ 0.033\end{tabular} & \begin{tabular}[c]{@{}l@{}}0.935\\ $\pm$ 0.048\end{tabular} & \begin{tabular}[c]{@{}l@{}}0.907\\ $\pm$ 0.059\end{tabular} & \begin{tabular}[c]{@{}l@{}}0.840\\ $\pm$ 0.080\end{tabular} & \begin{tabular}[c]{@{}l@{}}0.817\\ $\pm$ 0.090\end{tabular} \\ \cline{2-8} 
                             & Median Dice                                                    & \textbf{0.963}                                                    & 0.959                                                   & 0.948                                                   & 0.925                                                   & 0.861                                                   & 0.843                                                   \\ \hline
\end{tabular}
\label{table1}
\end{table*}
\begin{table*}
\caption{
DSC scores collected for the external testing data using both U-Net and DeepLab\textunderscore v3 in 512 space.}
\centering
\begin{tabular}{|c|c|l|l|l|l|l|l|}
\hline
\multicolumn{2}{|c|}{Image Resolution}                                                        & \multicolumn{1}{c|}{512$\times$512}                             & \multicolumn{1}{c|}{256$\times$256}                             & \multicolumn{1}{c|}{128$\times$128}                             & \multicolumn{1}{c|}{64$\times$64}                               & \multicolumn{1}{c|}{32$\times$32}                               & \multicolumn{1}{c|}{28$\times$28}                               \\ \hline
\multirow{2}{*}{U-Net}       & \begin{tabular}[c]{@{}c@{}}Mean Dice \\ $\pm$ Std Dev\end{tabular} & \begin{tabular}[c]{@{}l@{}}0.816 \\ $\pm$ 0.141\end{tabular} & \begin{tabular}[c]{@{}l@{}}0.899 \\ $\pm$ 0.063\end{tabular} & \begin{tabular}[c]{@{}l@{}}\textbf{0.902} \\ \textbf{$\pm$} \textbf{0.072}\end{tabular} & \begin{tabular}[c]{@{}l@{}}0.873 \\ $\pm$ 0.101\end{tabular} & \begin{tabular}[c]{@{}l@{}}0.845 \\ $\pm$ 0.086\end{tabular} & \begin{tabular}[c]{@{}l@{}}0.815 \\ $\pm$ 0.105\end{tabular} \\ \cline{2-8} 
                             & Median Dice                                                    & 0.853                                                    & 0.917                                                    & \textbf{0.918}                                                    & 0.897                                                    & 0.867                                                    & 0.838                                                    \\ \hline
\multirow{2}{*}{DeepLab\_v3} & \begin{tabular}[c]{@{}c@{}}Mean Dice \\ $\pm$ Std Dev\end{tabular} & \begin{tabular}[c]{@{}l@{}}0.918\\ $\pm$ 0.071\end{tabular}  & \begin{tabular}[c]{@{}l@{}}\textbf{0.922}\\ \textbf{$\pm$} \textbf{0.062}\end{tabular}  & \begin{tabular}[c]{@{}l@{}}0.911 \\ $\pm$ 0.068\end{tabular} & \begin{tabular}[c]{@{}l@{}}0.887\\ $\pm$ 0.087\end{tabular}  & \begin{tabular}[c]{@{}l@{}}0.812\\ $\pm$ 0.122\end{tabular}  & \begin{tabular}[c]{@{}l@{}}0.810\\ $\pm$ 0.088\end{tabular}  \\ \cline{2-8} 
                             & Median Dice                                                    & 0.931                                                    & \textbf{0.934}                                                    & 0.928                                                    & 0.906                                                    & 0.845                                                    & 0.834                                                    \\ \hline
\end{tabular}
\label{table2}
\end{table*}

\subsection{Experimental Design}
Our study was split into two distinct phases: manual detection and automatic detection. For our manual detection experimentation, 704 images were randomly chosen as testing images, while the remaining 98 images were used for validation. Additionally, 147 internal testing images were utilized alongside 385 external testing images from an independent cohort. On the other hand, for our automatic detection experimentation, 120 images, kept in their original resolutions ($>$ 1,000$\times$1,000 pixels), were used to evaluate the performance of our detect-then-segment framework relative to Mask-RCNN. 

The U-Net, DeepLab\textunderscore v3, and Mask-RCNN pipelines were deployed on a typical workstation with Intel Xeon CPU 2.2 GHz, 13 GB RAM, 33 GB Disk Space, 12 GB NVIDIA Tesla K80 GPU, and CUDA 10.1. For our manual detection experimentation, the hyper-parameters of the U-Net and DeepLab\textunderscore v3 pipelines were 150 epochs, a batch size of 4, a learning rate of 0.0001, a color space argument (RGB or LAB, depending on the trial), as well as a scale argument, which was altered to test the performance of segmentation across six distinct resolutions. Additionally, an Adam Optimizer was used to adaptively alter the learning rate, with beta values ranging from 0.9 to 0.999.

\subsubsection{Manual Detection}
Within our manual detection experimentation, manually detected glomerular images were processed with 2 distinct color spaces, 6 different image resolutions, and 2 unique segmentation backbones. We began with a U-Net segmentation framework, where the RGB and LAB color spaces were studied. Within each color space, 6 resolutions were tested through the U-Net pipeline: 1) 512$\times$512, 2) 256$\times$256, 3) 128$\times$128, 4) 64$\times$64, 5) 32$\times$32, and 6) 28$\times$28. For each trial, 150 epochs were run for all training, validation, and testing images. Additionally, within each epoch, a DSC value was generated for the validation and testing images. For each resolution, the epoch with the highest DSC value for the validation data was recorded and its generated model was saved. With this generated model, predicted masks were created at each of the six resolutions analyzed for the validation and testing data. This process was repeated yet again for the LAB color space. After this experiment was completed for the U-Net framework, the previously described methodology was also repeated for the DeepLab\textunderscore v3 framework, but the only color space analyzed for DeepLab\textunderscore v3 was the best performing color space in the U-Net trial. If the difference in performance between the two color spaces was negligible in the U-Net trial, then we defaulted to RGB to analyze the DeepLab\textunderscore v3 framework. This is because utilizing the RGB color space is standard, and the introduction of the LAB color space in our study was due to recent findings that show that the LAB color space produces better results for image classification tasks by reducing image channel-wise correlation~\cite{gowda2018colornet}.

Once all predicted masks were generated, the performance of the U-Net and DeepLab\textunderscore v3 segmentation networks was then analyzed by upsampling the predicted masks for the validation and testing images, and comparing them back to the original 512$\times$512 ground truth mask. Mean and median DSC scores were computed as a result of the upsampling.

\subsubsection{Automatic Detection}

In the automatic detection phase, the performance of our detect-then-segment framework was directly compared against Mask-RCNN through a cohort of 120 glomeruli images that were kept in their original resolution ($>$ 1000$\times$1000 pixels). In doing so, we utilized the model files that were generated for each of the six resolutions for both U-Net and DeepLab\textunderscore v3 as described in \textsection{4.2.1}. In particular, we first downsampled the 120 glomerular images to the scales of 512$\times$512, 256$\times$256, 128$\times$128, 64$\times$64, 32$\times$32, and 28$\times$28. Then, we procured each model file produced at the corresponding resolutions in the U-Net trial with the RGB color space. We then generated the predicted masks for each of the six downsampled resolutions of the 120 glomerular images utilizing the U-Net model file. We repeated this process for DeepLab\textunderscore v3 in the RGB color space. All predicted image sets were then upsampled back to the original size of the glomeruli images ($>$ 1,000$\times$1,000 pixels). Mean and median DSC values were then generated to evaluate performance. Furthermore, we also used a standard Mask-RCNN implementation to generate predicted masks at the corresponding original resolutions ($>$ 1,000$\times$1,000 pixels) of the glomeruli images. We then compared mean, median, and standard deviation DSC values to investigate which methodology performed best. 

\subsection{Evaluation Metrics and Statistical Methods}

DSC was the primary statistic used to evaluate segmentation performance. In particular, mean, median, and standard deviation of DSC were generated for analysis. Additionally, to evaluate statistical significance between each resolution and different methods, the Wilcoxon Rank Sum test was used with a significance threshold of either p $<$ 0.05 or p $<$ 0.01. A notched box plot was generated to visually demonstrate median DSC data, as well as the results of Wilcoxon Rank Sum test. Additionally, bar graphs were generated to show mean and standard deviation DSC data. Similarly, to demonstrate the relation between the performance of segmentation within each resolution in both LAB and RGB color space, DSC values for the U-Net trial were summarized in data tables.

\section{Results}

Our results presented in this section are divided into two central subsections to explore each aspect of our experimentation: 1) manual detection and 2) automatic detection.

\subsection{Manual Detection}

The first aspect of our study is manual detection, wherein we analyzed the key factors of segmentation, which include: segmentation backbones, input image resolutions, as well as color spaces. This phase of our study allowed us to better assess the conditions in which a detect-then-segment framework performs most optimally in the context of high resolution WSI.  

\begin{figure*}
\begin{center}
\includegraphics[width=1.0\textwidth]{{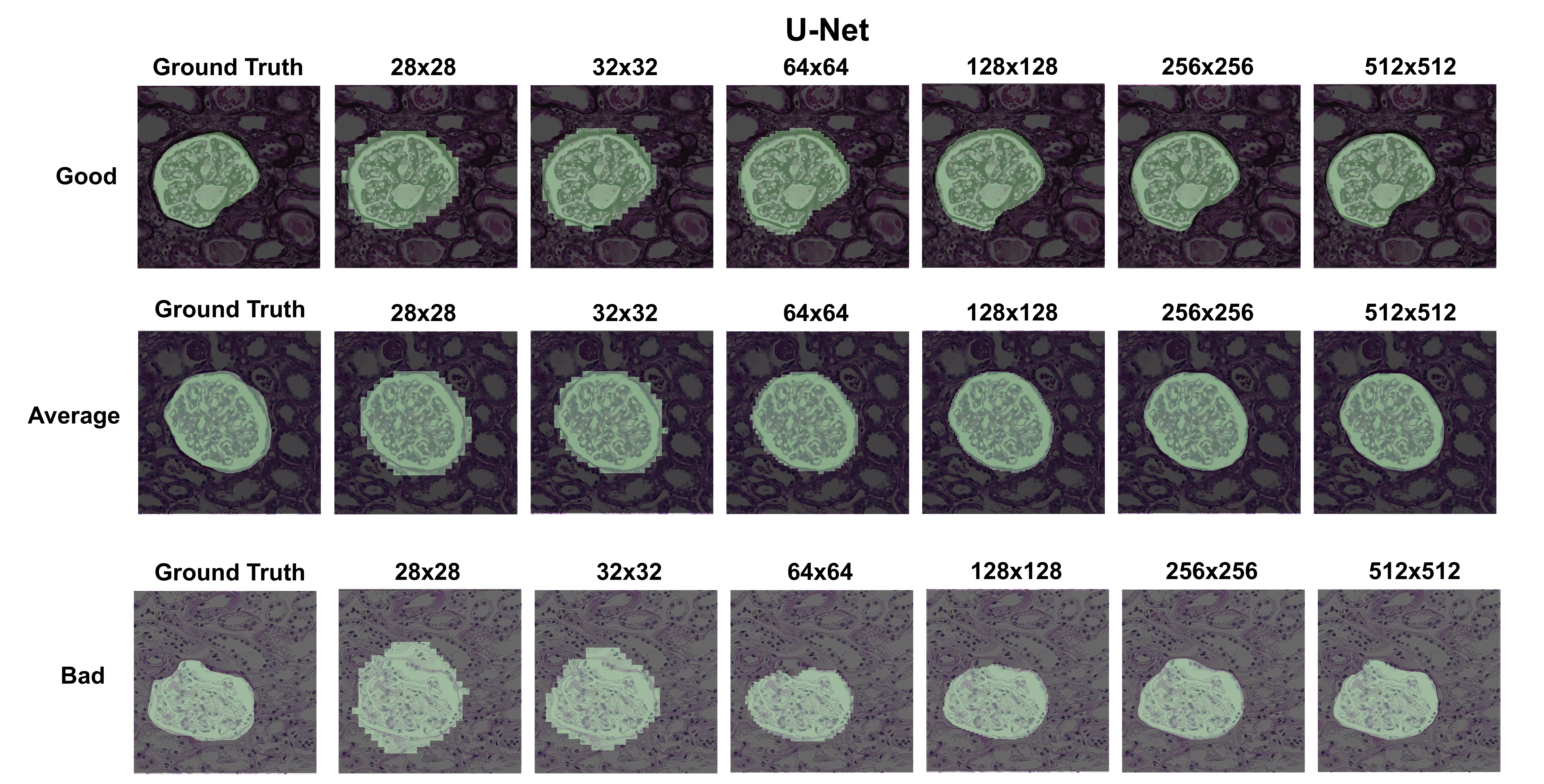}}
\end{center}
\caption{
   This figure demonstrates three categories of good, average, and bad performance which is defined by DSC. The background of each comparison is the ground truth input, and the overlaid image is the predicted mask. This specific figure shows the results of U-Net. 
} 
\label{Fig.3} 
\end{figure*}

\begin{figure*}
\begin{center}
\includegraphics[width=1.0\textwidth]{{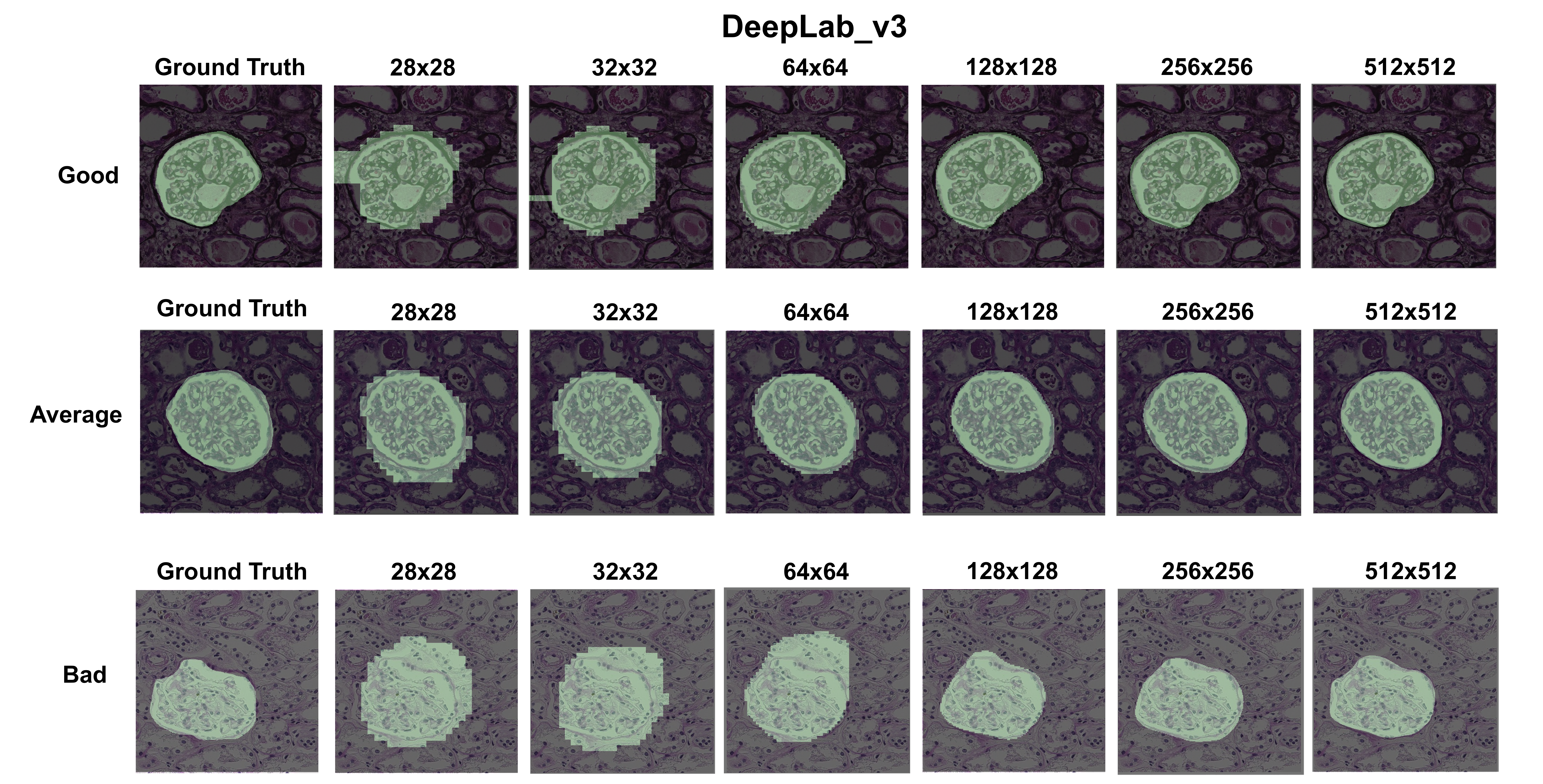}}
\end{center}
\caption{
       This figure demonstrates three categories of good, average, and bad performance which is defined by DSC. The background of each comparison is the ground truth input, and the overlaid image is the predicted mask. This specific figure shows the results of DeepLab\textunderscore v3. 
} 
\label{Fig.4} 
\end{figure*}

\begin{figure*}
\begin{center}
\includegraphics[width=1.0\textwidth]{{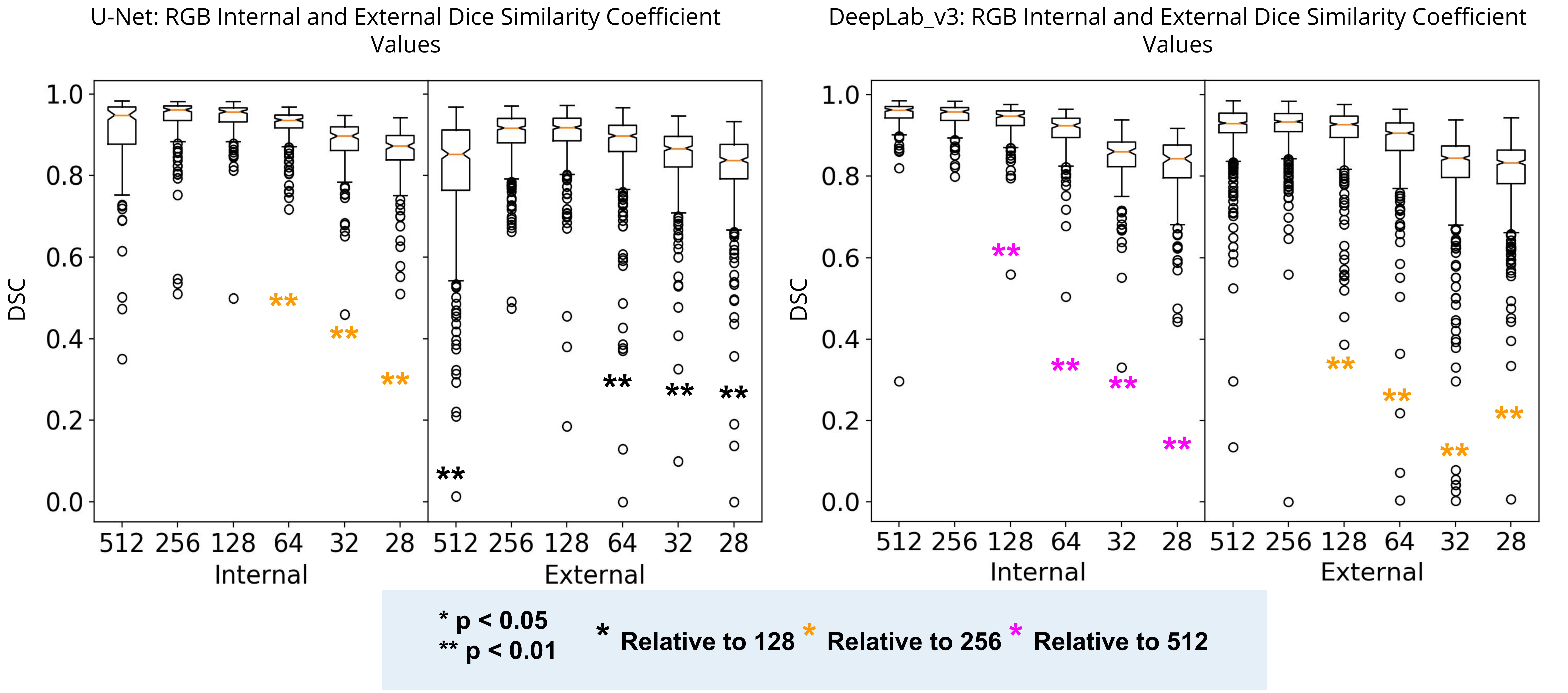}}
\end{center}
\caption{
     This figure demonstrates the notched box plots of each resolution for both internal and external testing data of the RGB color space. The left bar graph summarizes the results for the U-Net trial; the right graph summarizes the results for DeepLab\textunderscore v3. The legend and asterisks demonstrates results of computing the Wilcoxon Rank Sum Test on the specified resolutions. The median DSC values are derived by evaluating the ground truth mask in 512$\times$512 resolution, relative to the predicted mask in 512$\times$512 resolution for a fair comparison. 
} 
\label{Fig.5} 
\end{figure*}

\begin{figure*}
\begin{center}
\includegraphics[width=1.0\textwidth]{{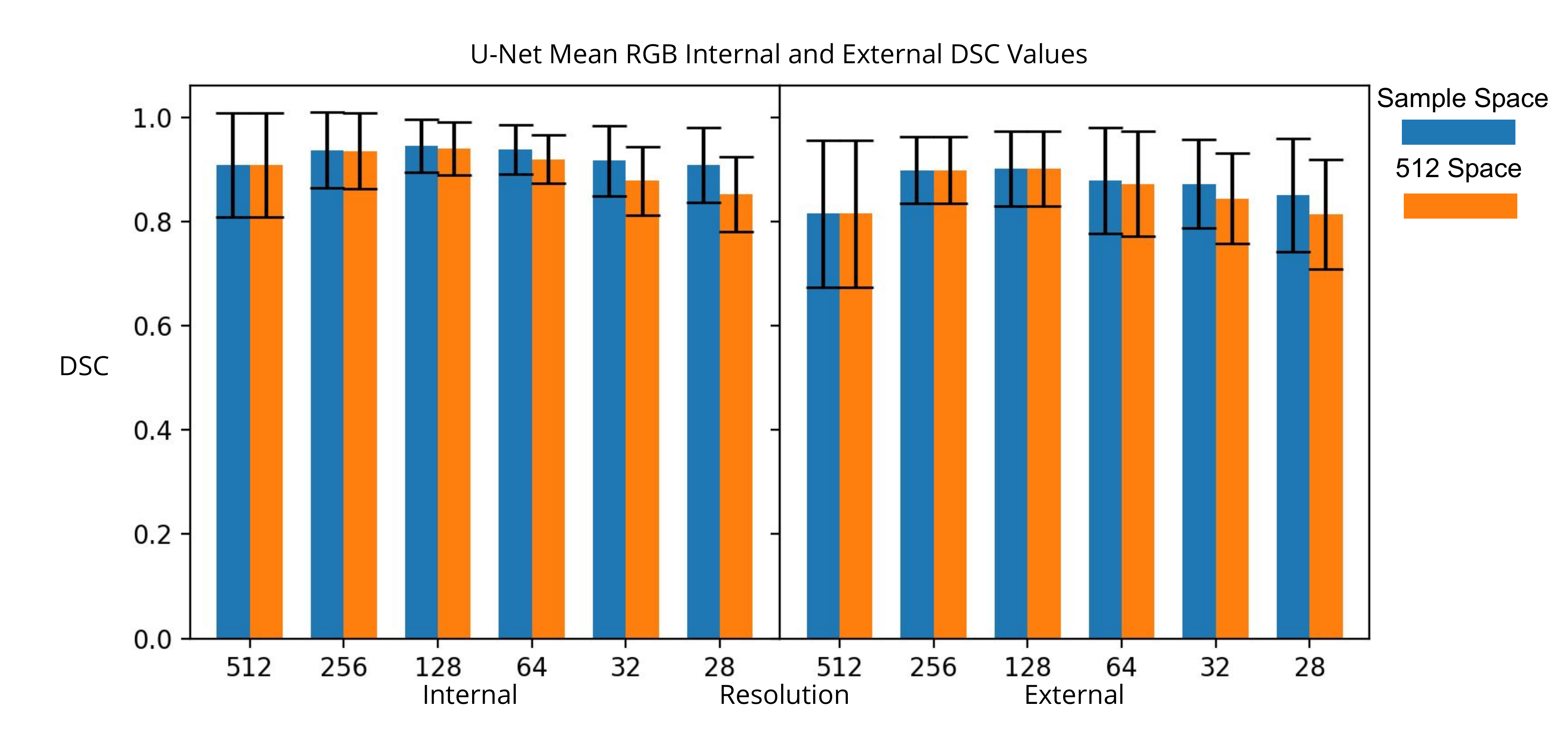}}
\end{center}
\caption{
     The above figure shows the mean and standard deviation for both internal and external testing images evaluated in both sample space and 512 space. Similar to how we define ``512 space", the term ``sample space" refers to the process by which the predicted masks that are produced across six distinct resolutions (512$\times$512, 256$\times$256, 128$\times$128, 64$\times$64, 32$\times$32, and 28$\times$28) are evaluated against the corresponding downsampled input image across the same six resolutions. We re-evaluated such images in 512 space for a fair DSC comparison through upsampling, as discussed earlier. As shown, 64$\times$64, 32$\times$32, and 28$\times$28 declined greatly in accuracy when comparing DSC in 512 space relative to DSC in sample space.
} 
\label{Fig.6} 
\end{figure*}

\begin{figure*}
\begin{center}
\includegraphics[width=1.0\textwidth]{{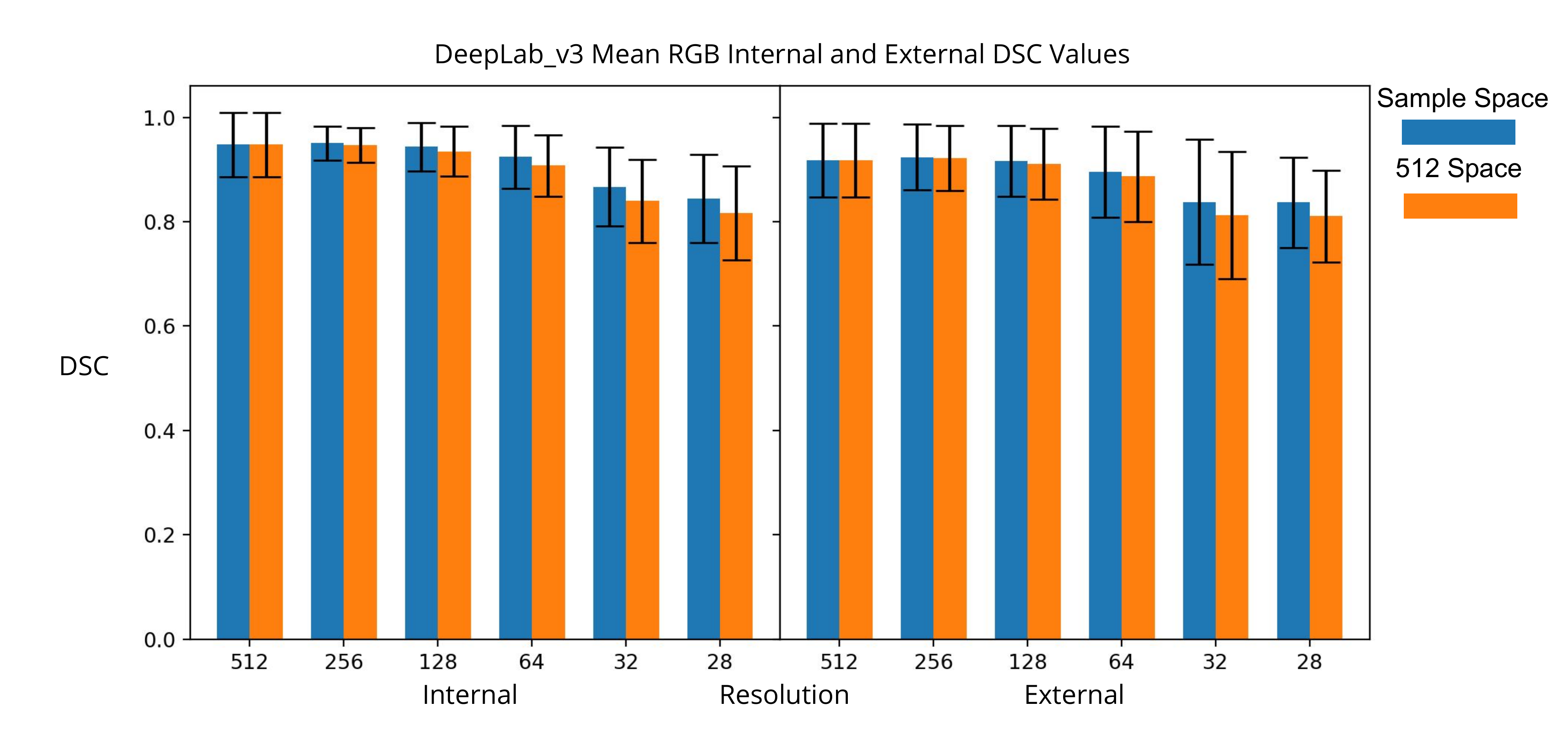}}
\end{center}
\caption{
The above figure shows the mean data and standard deviation for both internal and external testing images evaluated in both sample space and 512 space. As shown, 64$\times$64, 32$\times$32, and 28$\times$28 declined greatly in accuracy when comparing DSC in 512 space relative to DSC in sample space.
} 
\label{Fig.7} 
\end{figure*}

\begin{table*}
\caption{Performance of RGB and LAB color spaces on internal testing data using U-Net in 512 space.} 
\centering
\begin{tabular}{|c|l|l|l|l|l|l|l|}
\hline
\multicolumn{2}{|c|}{{\color[HTML]{000000} \textbf{Image Resolution}}}                                                                                 & {\color[HTML]{000000} 512$\times$512}                                                  & {\color[HTML]{000000} 256$\times$256}                                                  & {\color[HTML]{000000} 128$\times$128}                                                  & {\color[HTML]{000000} 64$\times$64}                                                    & {\color[HTML]{000000} 32$\times$32}                                                    & {\color[HTML]{000000} 28$\times$28}                                                    \\ \hline
{\color[HTML]{000000} }                                        & {\color[HTML]{000000} \begin{tabular}[c]{@{}l@{}}Mean Dice \\ $\pm$ Std Dev\end{tabular}} & {\color[HTML]{000000} \begin{tabular}[c]{@{}l@{}}0.909 \\ $\pm$ 0.099\end{tabular}} & {\color[HTML]{000000} \begin{tabular}[c]{@{}l@{}}0.936 \\ $\pm$ 0.073\end{tabular}} & {\color[HTML]{000000} \begin{tabular}[c]{@{}l@{}}\textbf{0.940} \\ \textbf{$\pm$ 0.051}\end{tabular}} & {\color[HTML]{000000} \begin{tabular}[c]{@{}l@{}}0.920 \\ $\pm$ 0.047\end{tabular}} & {\color[HTML]{000000} \begin{tabular}[c]{@{}l@{}}0.878 \\ $\pm$ 0.066\end{tabular}} & {\color[HTML]{000000} \begin{tabular}[c]{@{}l@{}}0.853 \\ $\pm$ 0.072\end{tabular}} \\ \cline{2-8} 
\multirow{-2}{*}{{\color[HTML]{000000} \textbf{RGB}}}    & {\color[HTML]{000000} Median Dice}                                                    & {\color[HTML]{000000} 0.948}                                                    & {\color[HTML]{000000} \textbf{0.961}}                                                    & {\color[HTML]{000000} 0.957}                                                    & {\color[HTML]{000000} 0.936}                                                    & {\color[HTML]{000000} 0.897}                                                    & {\color[HTML]{000000} 0.872}                                                    \\ \hline

{\color[HTML]{000000} }                                        & {\color[HTML]{000000} \begin{tabular}[c]{@{}l@{}}Mean Dice \\ $\pm$ Std Dev\end{tabular}} & {\color[HTML]{000000} \begin{tabular}[c]{@{}l@{}}0.917 \\ $\pm$ 0.081\end{tabular}} & {\color[HTML]{000000} \begin{tabular}[c]{@{}l@{}}0.936 \\ $\pm$ 0.068\end{tabular}} & {\color[HTML]{000000} \begin{tabular}[c]{@{}l@{}}\textbf{0.940} \\ \textbf{$\pm$ 0.050}\end{tabular}} & {\color[HTML]{000000} \begin{tabular}[c]{@{}l@{}}0.925 \\ $\pm$ 0.045\end{tabular}} & {\color[HTML]{000000} \begin{tabular}[c]{@{}l@{}}0.872 \\ $\pm$ 0.069\end{tabular}} & {\color[HTML]{000000} \begin{tabular}[c]{@{}l@{}}0.863 \\ $\pm$ 0.066\end{tabular}} \\ \cline{2-8} 
\multirow{-2}{*}{{\color[HTML]{000000} \textbf{LAB}}}    & {\color[HTML]{000000} Median Dice}                                                    & {\color[HTML]{000000} 0.946}                                                    & {\color[HTML]{000000} \textbf{0.961}}                                                    & {\color[HTML]{000000} 0.957}                                                    & {\color[HTML]{000000} 0.941}                                                    & {\color[HTML]{000000} 0.896}                                                    & {\color[HTML]{000000} 0.882}                                                    \\ \hline

\end{tabular}
\label{table3}
\end{table*}

\subsubsection{U-Net VS DeepLab\textunderscore v3}

We first present that both U-Net and DeepLab\textunderscore v3 confer particular advantages over one another across the six tested resolutions in the context of glomerular image data. {\color{green} \Tab~\ref{table1} and \ref{table2}} present DSC values for internal and external data for U-Net and DeepLab\textunderscore v3 in the RGB color space. Both tables show us that DeepLab\textunderscore v3 would perform better than U-Net for larger resolutions, such as 512$\times$512, 256$\times$256, and 128$\times$128, but would under-perform relative to U-Net for smaller image resolutions, such as 64$\times$64, 32$\times$32, and 28$\times$28. As shown, there is no clear, consistent framework that achieved the best DSC results for all trials. However, both U-Net and DeepLab\textunderscore v3 show distinct advantages --- DeepLab\textunderscore v3 tends to perform better for larger resolutions, whereas U-Net confers higher DSC values for smaller resolutions. 
\subsubsection{Image Resolution}

In the trial utilizing a U-Net framework, the resolution in RGB space with the highest median DSC value of 0.961 was 256$\times$256, which was statistically different relative to 64$\times$64, 32$\times$32, and 28$\times$28. On the other hand, for the external testing dataset in RGB space, the resolution with the highest median DSC value of 0.918 was 128$\times$128, which was significantly different relative to 28$\times$28, 32$\times$32, 64$\times$64, and 512$\times$512. A similar analysis was performed on the mean values and standard deviation of the internal and external testing DSC data of the RGB color space in the U-Net trial. For both internal and external data, 128$\times$128 was the resolution with highest mean DSC values in both the sample space and the upscaled 512 space. Additionally, {\color{red} \Fig~\ref{Fig.6}} further shows the resolutions of 64$\times$64, 32$\times$32, and 28$\times$28 experience the greatest decline in performance when comparing the DSC in sample space to the DSC in 512 space.

Considering the trial that used a DeepLab\textunderscore v3 framework, the highest median DSC value for internal data was 0.963, which occurred in the 512$\times$512 resolution. This particular resolution was significantly greater relative to 128$\times$128, 64$\times$64, 32$\times$32, and 28$\times$28. Similarly, for the external data, the highest median DSC value was 0.934 which occurred in 256$\times$256 space. This resolution was statistically greater relative to the resolutions of 128$\times$128, 64$\times$64, 32$\times$32, and 28$\times$28. Analyzing the mean data for DeepLab\textunderscore v3, it is clear that the highest mean DSC in the internal data occurred in the 256$\times$256 resolution when evaluating DSC in the sample space, whereas the highest mean DSC in 512 space occurred for the 512$\times$512 resolution.  For the external data, the highest mean DSC value occurred in the 256$\times$256 resolution when evaluating dice in the sample and 512 space. Similar to U-Net, the highest difference in mean DSC between the sample space and 512 space occurred in the resolutions of 64$\times$64, 32$\times$32, and 28$\times$28. The comparison of mean data can be seen visually in {\color{red} \Fig~\ref{Fig.7}}.

\begin{table*}
\caption{Results of U-Net and DeepLab\textunderscore v3 models during our automatic detection phase, as well as Mask-RCNN.}
\centering
\begin{tabular}{cllllllll}
\cline{1-7} \cline{9-9}
\multicolumn{1}{|c|}{Resolution}                                                                   & \multicolumn{1}{c|}{512$\times$512}                                                  & \multicolumn{1}{c|}{256$\times$256}                                                  & \multicolumn{1}{c|}{128$\times$128}                                                  & \multicolumn{1}{c|}{64$\times$64}                                                    & \multicolumn{1}{c|}{32$\times$32}                                                    & \multicolumn{1}{c|}{28$\times$28}                                                    & \multicolumn{1}{l|}{} & \multicolumn{1}{c|}{\begin{tabular}[c]{@{}c@{}}Mask \\ RCNN\end{tabular}}     \\ \cline{1-7} \cline{9-9} 
\multicolumn{1}{|c|}{\begin{tabular}[c]{@{}c@{}}\textbf{U-Net:}\\Mean Dice $\pm$ \\ Std Dev\end{tabular}}      & \multicolumn{1}{l|}{\begin{tabular}[c]{@{}l@{}}0.935 $\pm$ \\ 0.062\end{tabular}} & \multicolumn{1}{l|}{\begin{tabular}[c]{@{}l@{}}\textbf{0.947 $\pm$} \\ \textbf{0.037}\end{tabular}} & \multicolumn{1}{l|}{\begin{tabular}[c]{@{}l@{}}0.935 $\pm$ \\ 0.035\end{tabular}} & \multicolumn{1}{l|}{\begin{tabular}[c]{@{}l@{}}0.903 $\pm$ \\ 0.041\end{tabular}} & \multicolumn{1}{l|}{\begin{tabular}[c]{@{}l@{}}0.833 $\pm$\\ 0.060\end{tabular}}  & \multicolumn{1}{l|}{\begin{tabular}[c]{@{}l@{}}0.791 $\pm$ \\ 0.067\end{tabular}} & \multicolumn{1}{l|}{} & \multicolumn{1}{l|}{\begin{tabular}[c]{@{}l@{}}0.902 $\pm$ \\ 0.038\end{tabular}} \\ \cline{1-7} \cline{9-9} 
\multicolumn{1}{|c|}{\begin{tabular}[c]{@{}c@{}}\textbf{U-Net:}\\ Median Dice\end{tabular}}                 & \multicolumn{1}{l|}{0.956}                                                    & \multicolumn{1}{l|}{\textbf{0.957}}                                                    & \multicolumn{1}{l|}{0.945}                                                    & \multicolumn{1}{l|}{0.914}                                                    & \multicolumn{1}{l|}{0.845}                                                    & \multicolumn{1}{l|}{0.798}                                                    & \multicolumn{1}{l|}{} & \multicolumn{1}{l|}{0.908}                                                    \\ \cline{1-7} \cline{9-9} 
\multicolumn{1}{l}{}                                                                               &                                                                               &                                                                               &                                                                               &                                                                               &                                                                               &                                                                               &                       &                                                                               \\ \cline{1-7} \cline{9-9} 
\multicolumn{1}{|c|}{\begin{tabular}[c]{@{}c@{}}\textbf{DeepLab\_v3:}\\ Mean Dice $\pm$\\ Std Dev\end{tabular}} & \multicolumn{1}{l|}{\begin{tabular}[c]{@{}l@{}}\textbf{0.953 $\pm$} \\ \textbf{0.027}\end{tabular}} & \multicolumn{1}{l|}{\begin{tabular}[c]{@{}l@{}}0.941 $\pm$ \\ 0.034\end{tabular}} & \multicolumn{1}{l|}{\begin{tabular}[c]{@{}l@{}}0.919 $\pm$\\ 0.043\end{tabular}}  & \multicolumn{1}{l|}{\begin{tabular}[c]{@{}l@{}}0.876 $\pm$ \\ 0.053\end{tabular}} & \multicolumn{1}{l|}{\begin{tabular}[c]{@{}l@{}}0.771 $\pm$ \\ 0.067\end{tabular}} & \multicolumn{1}{l|}{\begin{tabular}[c]{@{}l@{}}0.750 $\pm$ \\ 0.067\end{tabular}} & \multicolumn{1}{l|}{} & \multicolumn{1}{l|}{\begin{tabular}[c]{@{}l@{}}0.902 $\pm$ \\ 0.038\end{tabular}} \\ \cline{1-7} \cline{9-9} 
\multicolumn{1}{|c|}{\begin{tabular}[c]{@{}c@{}}\textbf{DeepLab\_v3:}\\ Median Dice\end{tabular}}           & \multicolumn{1}{l|}{\textbf{0.961}}                                                    & \multicolumn{1}{l|}{0.950}                                                    & \multicolumn{1}{l|}{0.931}                                                    & \multicolumn{1}{l|}{0.891}                                                    & \multicolumn{1}{l|}{0.782}                                                    & \multicolumn{1}{l|}{0.753}                                                    & \multicolumn{1}{l|}{} & \multicolumn{1}{l|}{0.908}                                                    \\ \cline{1-7} \cline{9-9} 
\end{tabular}
\label{table4}
\end{table*}
The trend in the median DSC data and the results of the Wilcoxon Rank Sum Test can be seen in {\color{red} \Fig~\ref{Fig.5}}.

\subsubsection{Image Color Space}
   
The conferred accuracy of using either the LAB or RGB color spaces were found to be negligible in the trial utilizing a U-Net framework. {\color{green} \Tab~\ref{table3}} shows the results of the RGB and LAB color spaces on the internal dataset using U-Net. Both RGB and LAB give almost the same best DSC values, which is bolded. Therefore, the rest of the analysis in this paper is focused on the effectiveness of segmentation in the RGB color space, as the results are generalizable due to the similarity of segmentation performance between the RGB and LAB color spaces. In particular, the DeepLab\textunderscore v3 trial, as stated in the methodology, only performs its segmentation in the RGB color space, due to the results of U-Net. 

\subsection{Automatic Detection}

\subsubsection{Detect-then-Segment Framework VS End-to-End Mask-RCNN}
To begin, {\color{green} \Tab~\ref{table4}} demonstrates the results of applying our U-Net and DeepLab\textunderscore v3 models on the cohort of 120 images during our automatic detection phase, and generating predicted masks by downsampling the input images to six distinct resolutions. By evaluating the difference in performance between our proposed detect-then-segment pipeline relative to the standard end-to-end Mask-RCNN framework, we found that both U-Net and DeepLab\textunderscore v3 showed better mean and median DSC values for the resolutions of 512$\times$512, 256$\times$256, and 128$\times$128 (p $<$ 0.01, Wilcoxon Rank Sum). In particular, at best, our framework provides a mean DSC value of 0.953 via the DeepLab\textunderscore v3 backbone operating on a previously detected glomerulus of 512$\times$512 resolution, whereas Mask-RCNN produced a mean DSC value of 0.902. 

\section{Discussion}
First, in our experimentation with manual detection, we comprehensively searched and tested for the best detect-then-segment strategy. The results demonstrate that: 1) utilizing a higher resolution does not necessarily confer the best segmentation results; 2) the resolutions of 128$\times$128 and 256$\times$256 consistently demonstrated the best segmentation results; 3) lower resolutions (64$\times$64, 32$\times$32, and 28$\times$28) experience the greatest loss of accuracy when comparing DSC in sample space relative to 512 space; 4) DeepLab\textunderscore v3 yields better results in higher resolutions (512$\times$512, 256$\times$256, and 128$\times$128), whereas U-Net performs most optimally in lower resolutions (64$\times$64, 32$\times$32, and 28$\times$28); and 5) neither the LAB nor RGB color space give rise to better segmentation results relative to one another. Briefly, our results in \textsection{5.1.2} demonstrate that the resolutions of 128$\times$128 and 256$\times$256 consistently demonstrated the best DSC results, and the particular resolution of 512$\times$512 would actually yield relatively lower segmentation results, especially for median DSC. Additionally, the lower resolutions --- namely 64$\times$64, 32$\times$32, and 28$\times$28 --- consistently experienced a great loss in accuracy when analyzing the effectiveness of segmentation in 512 space. Finally, the results demonstrate that DeepLab\textunderscore v3 and U-Net perform most optimally in different ranges of resolutions. When considering RGB and LAB color spaces, we found there was no discernible effect or advantage of color space on the segmentation of high resolution glomerular images.

Further, we show that our proposed detect-then-segment pipeline is superior to the conventional end-to-end Mask-RCNN framework. Our summarized results show that the image resolutions of 512$\times$512, 256$\times$256, and 128$\times$128, in both U-Net and DeepLab\textunderscore v3 in RGB space are significantly better than that of Mask-RCNN. Of course, the most optimal result was achieved through DeepLab\textunderscore v3, with a mean DSC of 0.953 which occured in 512$\times$512 space. However, both U-Net and DeepLab\textunderscore v3 showed the same trend in the data. Overall, through our automatic detection trial, we conclude that utilizing a detect-then-segment framework across 512$\times$512, 256$\times$256, or 128$\times$128 will provide better segmentation results compared to the typical Mask-RCNN pipeline. Additionally, through our manual detection trial, we demonstrate that the most optimal detect-then-segment strategy involves utilizing a DeepLab\textunderscore v3 framework on larger resolution input images (256$\times$256 and 128$\times$128), in either the RGB or Lab color space. 

One key advantage of our research is our analysis of several critical factors of segmentation. In particular, our efforts strive to understand what works best for high resolution renal WSI, as opposed to practicing the standard end-to-end methods that are popular in computer vision. By studying the effect of color space, resolution, and segmentation backbone on the characterization of WSI through glomeruli data, we are better able to understand how to improve current segmentation networks that operate on high resolution images so as to yield better results. Another key advantage is how our study definitively shows that our proposed framework can yield a clear advantage in accuracy over the standard end-to-end instance segmentation methods in the context of high resolution renal WSI. Overall, our data provides a unique and important view towards new methodologies that show better results in high resolution imaging.

However, there are some important limitations to our study. First, the focus of this study was in the context of renal pathology and glomerular data. However, we expect the findings will be generalizable for other objects in renal pathology as the scaling issues are similar. Another limitation includes the fundamental restraints of the GPU when processing large scale images. In our study, the largest resolution photo on which we trained our data was 512$\times$512. In particular, we had to downsample the original resolution of the glomerular data due to the memory limitations of the GPU in order to efficiently conduct our study. 

A central way by which this study may be improved is by diversifying the dataset so as to include other biological objects, such as veins, arteries or tubules. Doing so would help solve the problem of a lack of generalizability with respect to our results, and would allow for the significance of our conclusions to be more widespread. 

\section{Conclusions}
Overall our experimentation through manual and automatic detection phases lead us to a three-fold conclusion: 1) a detect-then-segment framework is more effective than an end-to-end pipeline in the context of high resolution renal WSI; 2) the performance of a detect-then-segment framework is most optimal with a DeepLab\textunderscore v3 segmentation backbone operating on a 512$\times$512 resolution for previously detected glomerular input images; and 3) utilizing either RGB or LAB color spaces for previously detected glomerular input images does not yield a particular advantage over the other in a detect-then-segment framework. To conclude, our research paves the way towards further discussion and analysis in understanding effective and more nuanced methodologies that are more accurate than the current framework by which we characterize high resolution images of glomeruli, and biological objects of similar character, on large-scale WSI. 

\section*{Acknowledgment}
This work was supported in part by NIH NIDDK DK56942(ABF).

\ifCLASSOPTIONcaptionsoff
  \newpage
\fi
\bibliographystyle{IEEEtran}
\bibliography{main}
\end{document}